\documentclass{aa}

\usepackage{graphicx}
\usepackage{natbib}
\bibpunct{(}{)}{;}{a}{}{,}
\usepackage{txfonts}
\usepackage{color}

\newcommand{\ei}{\eta_I}
\newcommand{\eq}{\eta_Q}
\newcommand{\eu}{\eta_U}
\newcommand{\ev}{\eta_V}
\newcommand{\rqq}{\rho_Q}
\newcommand{\ru}{\rho_U}
\newcommand{\rv}{\rho_V}

\begin{document}
  \title{The usefulness of analytic response functions}

  \author{D. Orozco Su\'arez \inst{1} \and J.C. del Toro Iniesta \inst{1}}
    \offprints{D. Orozco Su\'arez} \institute{Instituto de Astrof\'isica de
    Andaluc\'ia, Consejo Superior de Investigaciones Cient\'ificas, Apdo. de
    Correos 3004, E-18080 Granada, Spain \\ \email{orozco@iaa.es;jti@iaa.es} }

    \date{Received August 7, accepted October 8, 2006}

    \abstract {}{We introduce analytical response
	functions and their main properties as an important diagnostic tool
	that help understand Stokes profile formation physics and
	the meaning of well-known behaviors of standard inversion codes of the
	radiative transfer equation often used to measure solar magnetic fields.} {A
	Milne-Eddington model atmosphere is used as an example where
	response functions are analytical. A sample spectral line has been
	chosen to show the main qualitative properties.}{We show that analytic
	response functions readily provide explanations for various well-known
	behaviors of spectral lines, such as the sensitivity of visible lines to
	weak magnetic fields or the trade-offs often detected in inversion codes
	between the Milne-Eddington thermodynamic parameters. We also show that
	response functions are helpful in selecting sample wavelengths
	optimized for specific parameter diagnostics.}{}

\maketitle

\section{Introduction}
\label{secintro}

Diagnosing the solar atmosphere from spectropolarimetric observations
is one of the most important subjects of current solar physics. Both
the theoretical understanding of the physical processes taking place
in the photosphere and the design of new instrumentation that improve
our ability to obtain more and better information from the Sun can be
improved by a thorough study of the radiative transfer equation
(hereafter referred to as RTE). RTE is the only tool available to
describe the problem mathematically. Approximations have been devised
according to the observational and the {\em post-facto} computational
capabilities. The Milne-Eddington (M-E) approximation has provided
insight into the physical processes taking place in line formation and
inferring accurate values of several physical parameters of the solar
atmosphere. Its specific analytical character is its most powerful
feature.

A physical analysis of the sensitivities of spectral lines in terms of
analytic mathematical functions is still missing in the literature and
may provide a better understanding of how the solar parameters
influence the shape of the observed Stokes profiles of these spectral
lines and explanations for the trade-offs and other well known
behaviors of inversion codes currently used for the inference of such
solar atmospheric parameters. Here we introduce the {\em analytic}
response functions (RFs) of Stokes profiles as formed in M-E model
atmospheres and discuss their main properties.

Weighting functions for unpolarized light (Mein 1971) were the
precursors of RFs, extended to polarized light by Landi Degl'Innocenti
\& Landi Degl'Innocenti (1977). As explained by Ruiz Cobo \& del Toro
Iniesta (1994; see also del Toro Iniesta 2003), RFs provide the
sensitivities of Stokes profiles to the various atmospheric quantities
playing a role in line formation. Since all these quantities are
constant with depth in a M-E atmosphere, M-E RFs are simply partial
derivatives of the analytic solution of the RTE with respect to the
model parameters. This feature enables us to deduce analytic formulae
for the sensitivities (they are explicitly written in the Appendix)
and to study their characteristics and properties. Such properties are
shown to be useful in practice in understanding the behavior of
spectral lines as well as in helping in line and sample selection
when designing new instruments.

\section{The response functions in a Milne Eddington atmosphere}

\subsection{Summary of radiative transfer}

The radiative transfer equation (RTE) for polarized light in a
plane-parallel atmosphere is
\begin{equation}
\frac{d\vec{I}}{d \tau}=\mathbf{K}(\vec{I}-\vec{S}),
\end{equation}
where $\vec{I}=(I,Q,U,V)^\mathrm{T}$ stands for the Stokes vector which gives
a full description of the polarization state of light, $\tau$ for the
continuum optical depth at a reference wavelength, $\mathbf{K}$ for the 4x4
propagation matrix, $\vec{S}$ for the source function vector and $^\mathrm{T}$
means the transpose. All the medium properties relevant to line formation are
contained in $\mathbf{K}$ and $\vec{S}$. In LTE conditions,
$\vec{S}=(B_\lambda(T),0,0,0)^\mathrm{T}$, where $B_\lambda(T)$ is the Planck
function at the local temperature $T$.

In a Milne-Eddington (M-E) model atmosphere, an analytical solution is found for the
RTE \citep[see, e.g.][]{1956PASJ....8..108U,Rakk1,Rakk2,1982SoPh...78..355L}. In
such an atmosphere, all the atmospheric quantities are constant with depth
except for the source function that varies linearly:
\begin{equation}
\vec{S}=\mathbf{S}_0+\mathbf{S}_1\tau = (S_0 + S_1 \tau)(1,0,0,0)^\mathrm{T}.
\end{equation}

The propagation matrix is also constant with depth. Following, e.g., the
notation in \citet{2003insp.book.....D}, such an analytical solution reads
\begin{eqnarray}
\label{eqsolution}
I & = & S_0+ \Delta^{-1}[\ei(\ei^2+\rqq^2+\ru^2+\rv^2)]\, S_1,  \nonumber \\
Q & = & -\Delta^{-1}[\ei^2 \eq+\ei(\ev\ru-\eu\rv)+\rqq \Pi]\,S_1,  \nonumber \\
U & = & -\Delta^{-1}[\ei^2 \eu+\ei(\eq\rv-\ev\rqq)+\ru \Pi]\, S_1,  \nonumber \\
V & = & -\Delta^{-1}[\ei^2 \ev+\ei(\eu\rqq-\eq\ru)+\rv \Pi]\, S_1, 
\end{eqnarray}
with
\begin{equation}
\Delta = \ei^2(\ei^2-\eq^2-\eu^2-\ev^2+\rqq^2+\ru^2+\rv^2)-\Pi^2,
\end{equation}
where
\begin{equation}
\Pi = \eq\rqq+\eu\ru+\ev\rv.
\end{equation}

It can easily be seen that $\ei$, $\eq$, $\eu$, $\ev$, $\rqq$, $\ru$,
and $\rv$, and hence the solution depends on just nine parameters,
($B$, $\gamma$, $\chi$), the strength, inclination and azimuth of the
magnetic field vector on the local reference frame, on $S_0$, $S_1$,
the two parameters describing the source function, on $\eta_0$, the
line-to-continuum absorption coefficient ratio, on
$\Delta\lambda_\mathrm{D}$, the Doppler width of the line, on the
damping parameter $a$, and on the line-of-sight velocity,
$v_\mathrm{LOS}$.

\subsection{Milne-Eddington response functions}

According to \cite{1994A&A...283..129R} (see also
\citealt{1996SoPh..164..169D} or \citealt{2003insp.book.....D}) the
sensitivity of the Stokes profiles to perturbations of the atmospheric
physical quantities is given by the response functions
(RFs). Fortunately, in the specific case of constant quantities (model
parameters) with depth, as is the case of an M-E atmosphere, such RFs
are the result of integrating in depth the regular RFs. Such
\textit{$\tau$-integrated} response functions are thus simply
functions of the wavelength and can be considered as the partial
derivatives of the Stokes vector with respect to the corresponding
model parameter:
\begin{equation}
\label{eqresponse}
\vec{R}_{x} (\lambda) = \frac{\partial\vec{I}(\lambda)}{\partial x},
\end{equation}
where $x$ represents any of the model parameters. We hereafter
refer to these $\tau$-integrated RFs as M-E RFs or just RFs.

Therefore, by taking derivatives of the analytical
solution (\ref{eqsolution}), the sensitivities of the Stokes profiles to
perturbations of the M-E model parameters can be found (see
the Appendix for explicit formulae). These sensitivities are the
only tools we have to evaluate our ability of determining the various
quantities: should the $\vec{I}$ Stokes vector not vary after a perturbation
of a parameter, $x$, we would be unable to infer it from the observations (it
would not be a proper model parameter).

\begin{figure*}[!ht]
  \centering
\resizebox{0.9\hsize}{!}{\includegraphics[width=\textwidth]{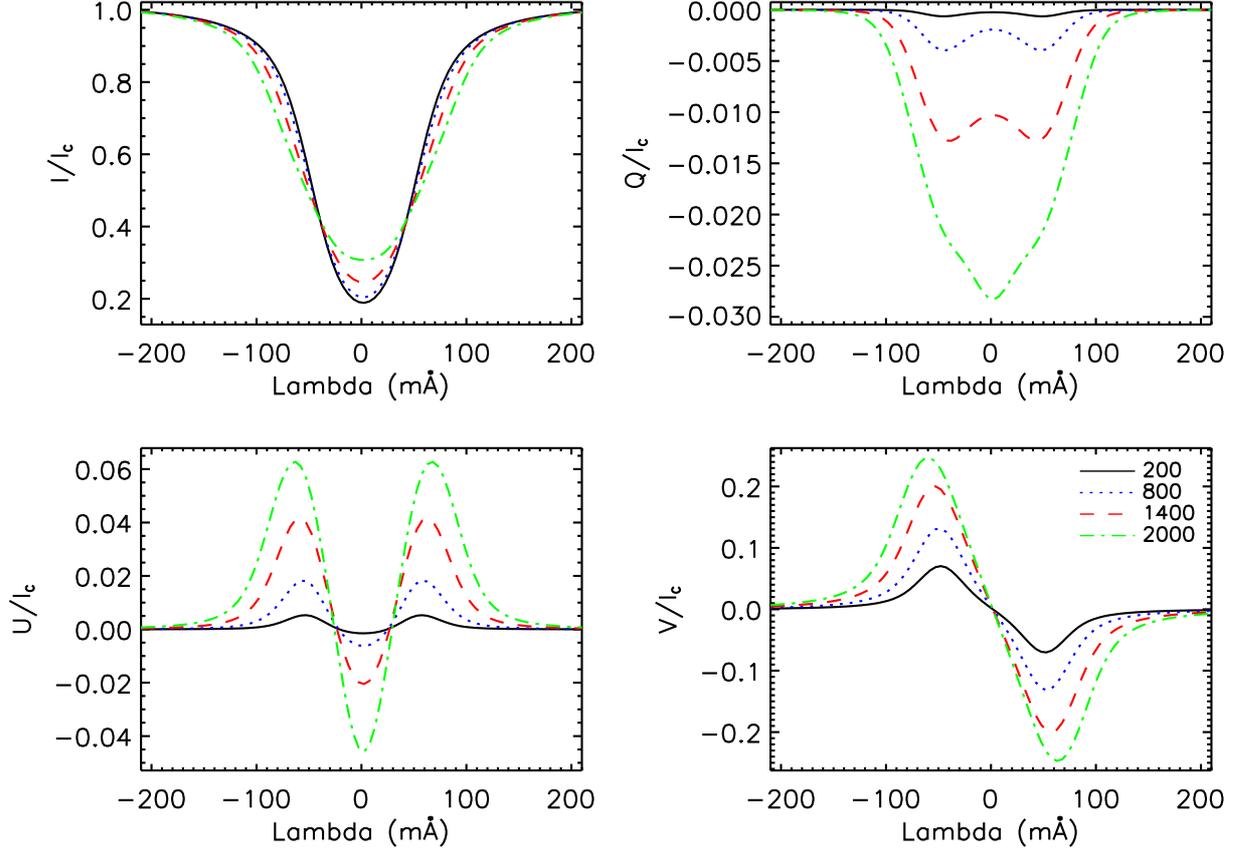}}
  \caption{Stokes $I/I_{\mathrm{c}}$, $Q/I_{\mathrm{c}}$,
  $U/I_{\mathrm{c}}$ and $V/I_{\mathrm{c}}$, for the \ion{Fe}{i} line at
  525.06 nm, with a magnetic inclination and azimuth of 45
  degrees. Different lines stand for different magnetic field strength
  values. The Stokes parameters are normalized to the local
  continuum.}
  \label{Figstokes}
\end{figure*}

\subsection{Line sensitivities: the shape of M-E RFs}

Equations (\ref{eqsolution}) and (\ref{eqresponse}) provide the
necessary means for studying the behavior of the M-E Stokes
profiles. The shapes of the RFs do not vary dramatically from model to
model or from line to line. M-E RFs appear homologous to each
other. This property allows us to choose a single line to illustrate
the practical usefulness of our functions. Let us take the \ion{Fe}{i}
line at 525.064 nm as an example. We select this line because it is
used by the IMaX magnetograph \citep{2004SPIE.5487.1152M} and some of
the results have implications either for the design or for the
analysis of the data to be obtained with this magnetograph. The line
has an effective Land\'e factor of 1.5 and is often considered to be
quite insensitive to temperature perturbations
\citep[e.g.,][]{1984A&A...131..333S}. A single model is also enough
for our purposes. We have used the NSO Fourier Transform Spectrometer
atlas as a reference spectrum and the line was fit with errors smaller
than a 2\%. The resulting model parameters are: $S_0=0.02$, $S_1=1$,
$\eta_0=7.2$, $a=0.3$, $\Delta\lambda_\mathrm{D}=30$ m\AA\/ and a
macroturbulent velocity, $v_\mathrm{mac}=0.37$ km/s. Unless otherwise
stated, all the numerical examples that follow refer to this line and
this model. Several magnetic field strengths (200, 800, 1400 and 2000
G) have been used to synthesize the Stokes profiles and their RFs,
assuming a constant field inclination and azimuth of 45 degrees.

\begin{figure*}[!t]
  \centering
  \resizebox{0.9\hsize}{!}{\includegraphics[width=\textwidth]{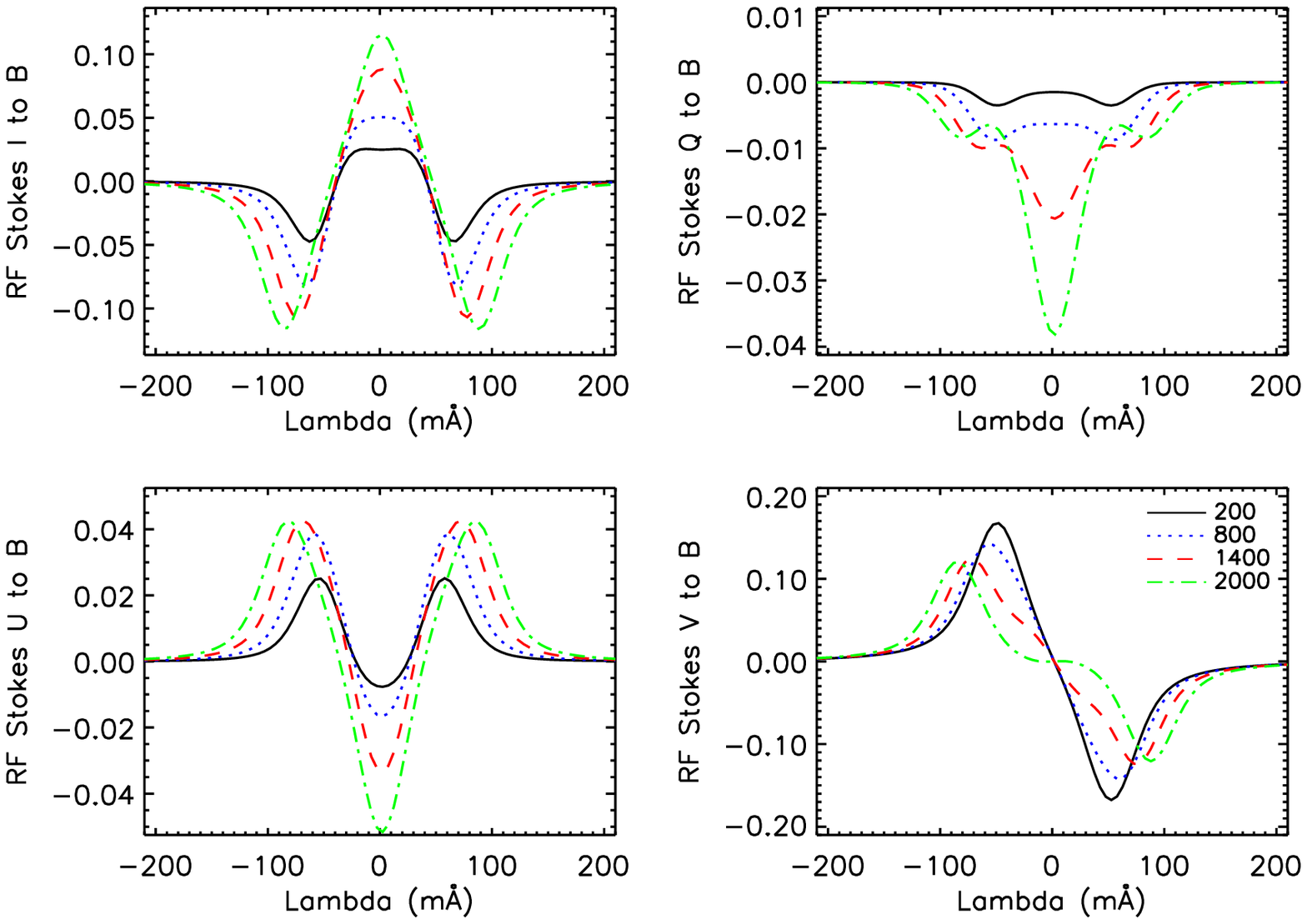}}
  \caption{Analytical M-E RFs of Stokes $I/I_{\mathrm{c}}$,
  $Q/I_{\mathrm{c}}$, $U/I_{\mathrm{c}}$ and $V/I_{\mathrm{c}}$ to
  magnetic field strength for the \ion{Fe}{i} line at 525.06 nm, with
  a magnetic inclination and azimuth of 45 degrees. Different lines
  stand for different magnetic field strength values. Units are
  10$^{-3}$ G$^{-1}$.}
  \label{Fig1rfs}
\end{figure*}

\begin{figure*}[!t]
  \centering
  \resizebox{0.9\hsize}{!}{\includegraphics[width=\textwidth]{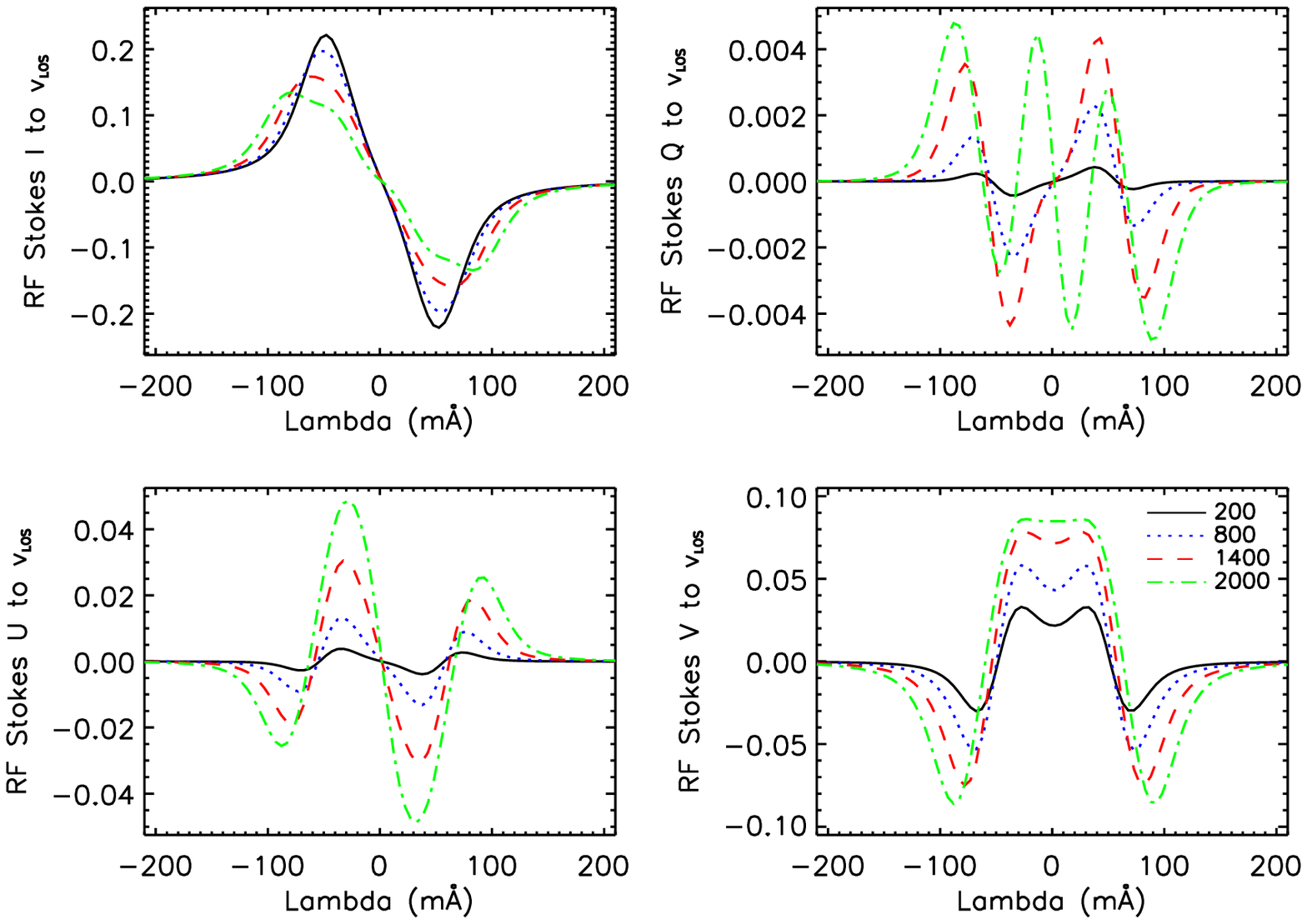}}
  \caption{Analytical M-E RFs of Stokes $I/I_{\mathrm{c}}$,
  $Q/I_{\mathrm{c}}$, $U/I_{\mathrm{c}}$ and $V/I_{\mathrm{c}}$ to LOS
  velocity for the \ion{Fe}{i} line at 525.06 nm, with a magnetic
  inclination and azimuth of 45 degrees. Different lines stand for
  different magnetic field strength values. Units are [km/s]$^{-1}$.}
  \label{Fig2rfs}
\end{figure*}

Fig.~\ref{Figstokes} shows the synthesized Stokes profiles. As the
magnetic field increases, the Stokes $V$ lobes increase but their
peaks do not separate much because the strong field regime has not
been reached for this line with these strengths. In
Fig.~\ref{Fig1rfs}, we give a graphical illustration of the analytical
RFs of the four Stokes parameters to magnetic field strength
perturbations. Both the Stokes profiles and the RFs present
wavelength symmetry properties, as expected from a M-E model
atmosphere. The RFs to the magnetic field strength preserve the Stokes
profile symmetries while velocity RFs display opposite parity (see
Fig.~\ref{Fig2rfs}).

Fig.~\ref{Fig1rfs} shows that the response of the line is wavelength
dependent. Different wavelength positions have different
sensitivities. Within a single Stokes profile different wavelength
samples react differently to the same perturbation. Some of the
samples are insensitive. For instance, in this example the Stokes $V$
zero-crossing point remains the same regardless of $B$ and, hence,
the response is zero at this wavelength. All the RFs show peaks
corresponding to different maxima and minima. Note that these extrema
pinpoint where the Stokes profiles are more sensitive to perturbations
of the physical quantity: the greater the peak, the greater the
sensitivity.

Although Stokes $I$, $Q$ and $U$ are more sensitive to $B$
perturbations when the strength is greater, the Stokes $V$ profile
sensitivity to field strength perturbations is a maximum for the weak
fields and decreases with increasing field strength. In the weak field
regime, Stokes $V$ is proportional to $B$ and any change of $B$ is
translated directly into an increase (or a decrease) of the $V$
signal; when the field increases, however, a competition between
increasing the profile and peak separation becomes important; At a
given $B$ value, peaks will no longer increase but separate from each
other. This behavior of Stokes profiles is known for long but a glance
to the Stokes $V$ panel of Fig. 2 illustrates it in a very clear
way. Moreover, the sensitivity of Stokes $V$ in the weak field regime
explains the reasonably accurate inversion results for weak magnetic
fields obtained in numerical experiments by
\cite{1998ApJ...494..453W}.

Fig.~\ref{Fig2rfs} shows the Stokes RFs to LOS velocity. The first clear
feature in this figure is that neither the sizes nor the shapes depend on the
LOS velocity. The latter only shifts the RFs as it does with the profiles. The
RF size is larger for Stokes $I$ and $V$ than for Stokes $Q$ and $U$,
because of the corresponding size of the profiles. Since Stokes $I$ and $V$
are larger than Stokes $Q$ and $U$ in this example, most information on
velocities is carried by $I$ and $V$. The LOS velocity can always be well
determined because the loss of sensitivity to $v_\mathrm{LOS}$ perturbations
of the Stokes $I$ profile is compensated by that of the $V$ profile.

 The Stokes $I$ RF to LOS velocity perturbations decreases with $B$ when the
 Stokes $Q$, $U$, and $V$ RFs increase. This is due to the
 different shape ratios of the various profiles. According to Cabrera Solana,
 Bellot Rubio and del Toro Iniesta (2005), the spectral line sensitivity to
 the LOS velocity is mostly determined by the ratio between the width and the
 depth of the line. The greater the field strength, the wider
 and shallower the Stokes $I$ profile. Therefore, its sensitivity to
 $v_\mathrm{LOS}$ perturbations decreases with increasing $B$. Each lobe of
 Stokes $V$, however, first becomes larger and then narrower and steeper at
 the central wavelength as $B$ increases. Hence its greater sensitivity to
 $v_\mathrm{LOS}$ for increasing field strengths.

The relative maxima of the RFs to LOS velocity perturbations correspond to
wavelength positions where the inflection points of the Stokes profiles are
located independently of the model atmosphere and spectral line. For
instance, the minimum of Stokes $I$ and the peaks of Stokes $V$ correspond to
zeros on the corresponding RFs to LOS velocity, therefore regions where the
Stokes profiles do not change when LOS velocity does.

\begin{figure*}[!t]
  \centering
  \resizebox{0.9\hsize}{!}{\includegraphics[width=\textwidth]{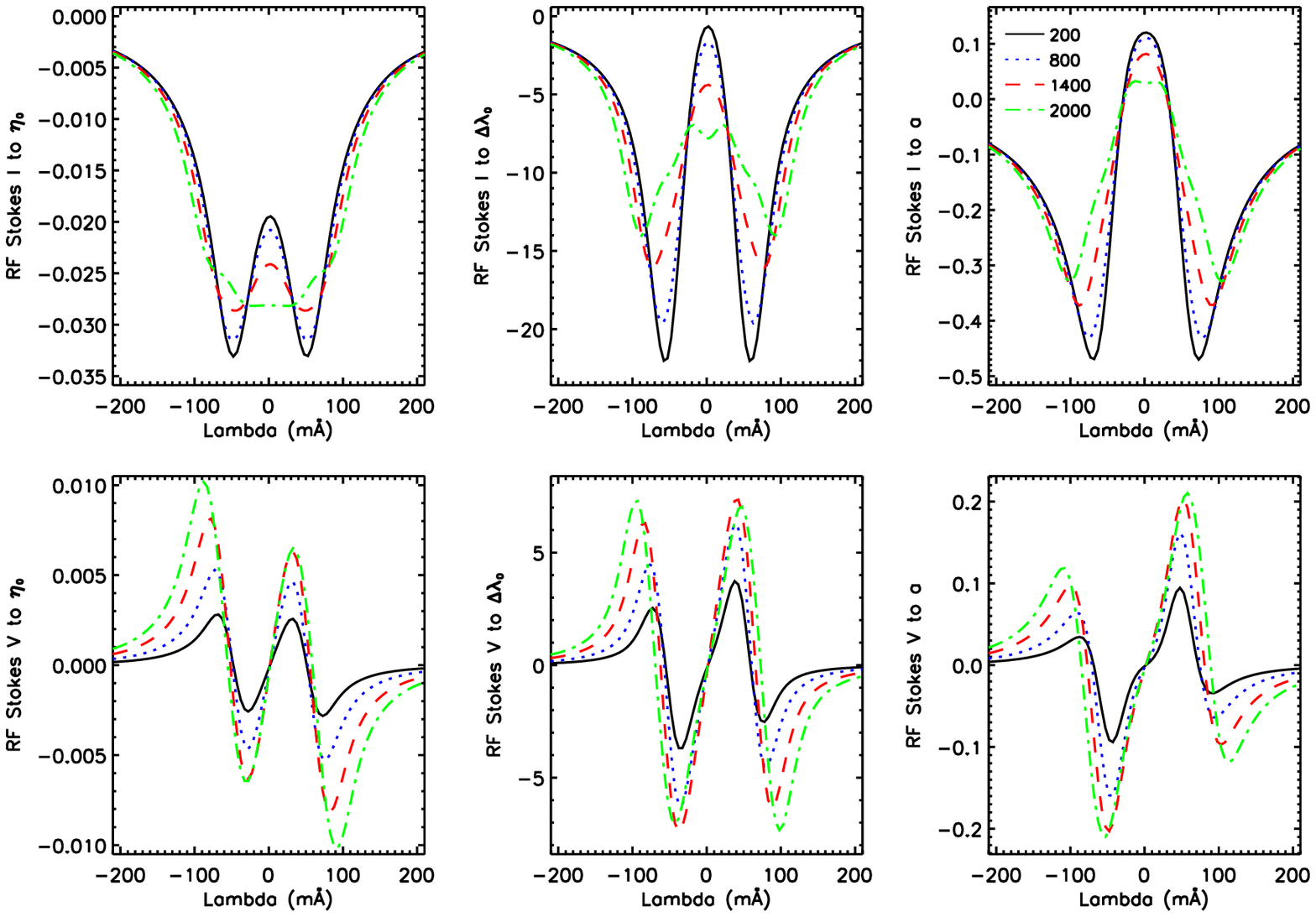}}
  \caption{Analytical M-E RFs of Stokes $I/I_{\mathrm{c}}$ (upper
  panels) and $V/I_{\mathrm{c}}$ (bottom panels) to $\eta_0$, to
  $\Delta\lambda_\mathrm{D}$ and to $a$ (left, middle and right panels
  respectively), for the \ion{Fe}{i} line at 525.06 nm, with a
  magnetic inclination and azimuth of 45 degrees. Different lines
  stand for different magnetic field strength values. Units are
  dimensionless for the left and right panels since $\eta_0$ and $a$
  are dimensionless. Units for the middle panels are \AA$^{-1}$. Note the
  similarities among the different RFs.}
  \label{Figeta}
\end{figure*}

The extrema of the RFs to $B$ and to $v_\mathrm{LOS}$ perturbations do
not coincide with those of the corresponding profiles. This fact can
be clearly seen in, e.g., the bottom right panels of
Figs.~\ref{Fig1rfs} and \ref{Fig2rfs}. Therefore, the extrema of the
Stokes profiles do not carry, in principle, more information on
given parameters than any other particular wavelength sample. Another
very interesting feature is that, for a given spectral line, the RFs
differ from each other. RFs to magnetic field strength perturbations
do not resemble those to LOS velocity perturbations (compare
Figs.~\ref{Fig1rfs} and \ref{Fig2rfs}). For instance, their maximum
sensitivities (RF peaks) are placed at different wavelengths. These
differences among RFs help disentangle the influences on spectral line
formation of the various model quantities and allow inversion
algorithms based on RFs \citep[e.g., SIR by][]{1992ApJ...398..375R} to
obtain accurate results: if a given Stokes profile is inappropriate for a
particular wavelength sample, other profile or wavelength samples
provide the required information. RF differences can also be seen for
the other M-E parameters except for $\Delta\lambda_\mathrm{D}$,
$\eta_0$ and $a$. The RFs to these thermodynamic parameter
perturbations are very similar to each other as can be seen in
Fig.~\ref{Figeta}. A small perturbation of any of these three
parameters produces a modification in the Stokes profiles that is very
similar to the changes produced by small perturbations of the other
two. These similarities between the $\Delta\lambda_\mathrm{D}$,
$\eta_0$ and $a$ RFs explain the trade-offs often observed in M-E
inversions. Fortunately, their RFs are different enough
from those of the other model parameters for them to be accurately
retrieved \citep[see, e.g.,][]{1998ApJ...494..453W}. Thus,
the M-E model atmosphere, although providing a
simplistic scenario for line formation which may not
full account for all thermodynamic properties, allows fairly accurate
inferences of the constant magnetic field vector $\vec{B}$,
$v_\mathrm{LOS}$, $S_0$ and $S_1$.

\begin{figure*}[!t]
  \centering
  \resizebox{0.9\hsize}{!}{\includegraphics[width=\textwidth]{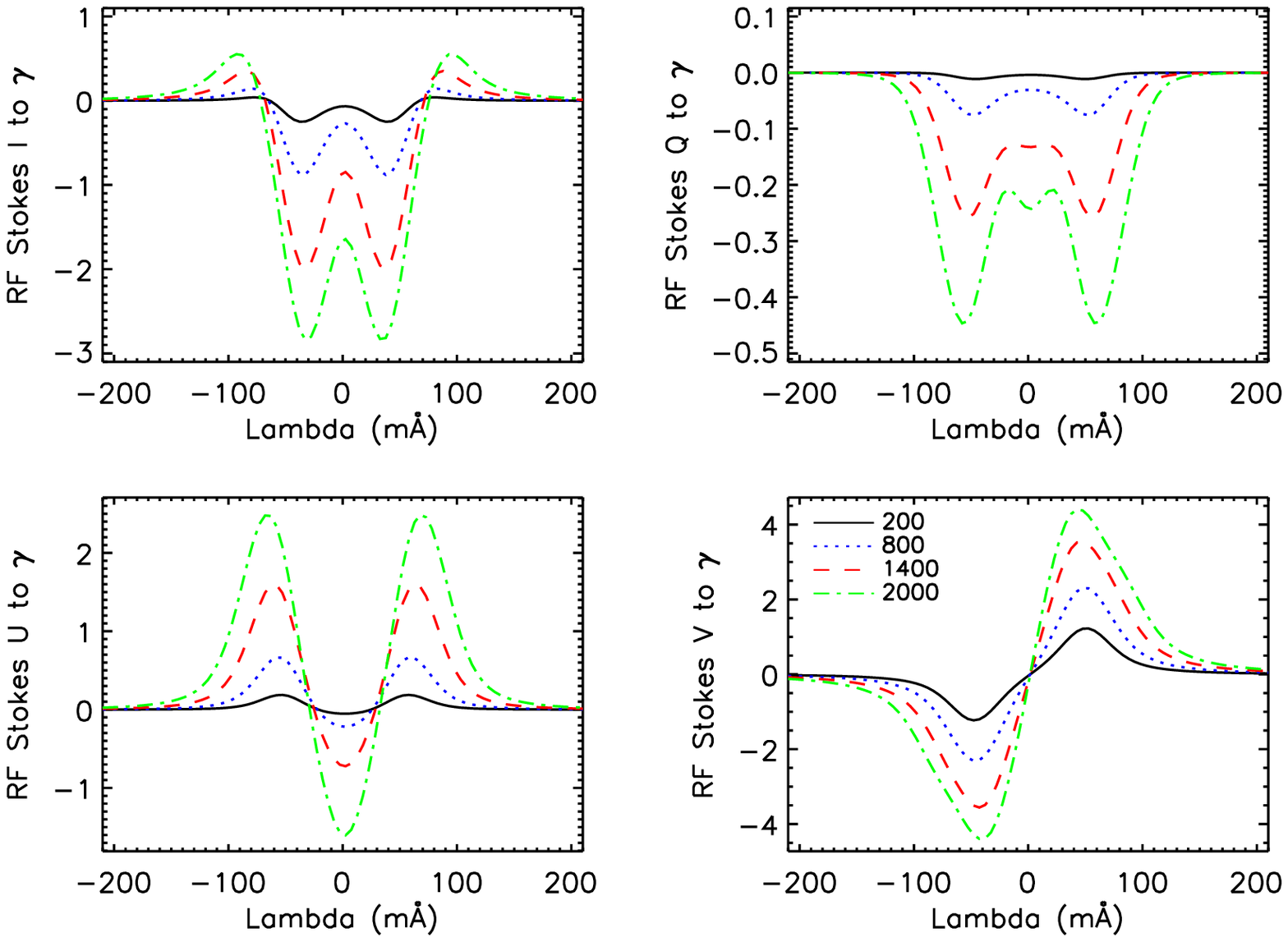}}
  \caption{Analytical M-E RFs of Stokes $I/I_{\mathrm{c}}$,
 $Q/I_{\mathrm{c}}$, $U/I_{\mathrm{c}}$ and $V/I_{\mathrm{c}}$ to
 magnetic field inclination, $\gamma$, for the \ion{Fe}{i} line at
 525.06 nm, with a magnetic inclination and azimuth of 45
 degrees. Different lines stand for different magnetic field strength
 values. Units are in 10$^{-3}$ [degrees]$^{-1}$.}
  \label{Fig3rfs}
\end{figure*}

The RFs to magnetic field inclination and azimuth perturbations do not
depend on the derivatives of the absorption and dispersion profiles;
thus, the shapes of the RFs are very similar to the corresponding
Stokes profiles (see Fig.~\ref{Fig3rfs}).  Only Stokes $Q$ and $U$
respond to azimuth perturbations. The larger the field strength, the
greater the sensitivity of the Stokes profiles to $\gamma$ and $\chi$
perturbations. This is again an explanation of a well known fact: we
measure $\gamma$ and $\chi$ more accurately when $B$ is strong.

\subsection{Relative response functions}

So far we have only discussed ``absolute'' RFs, i.e., functions with
dimensions; e.g. the RF to $B$ is measured in G$^{-1}$, that to
$v_\mathrm{LOS}$ perturbation is measured in (km s$^{-1}$)$^{-1}$ and
so on: RFs give modifications of the profile per unit perturbation of
the parameter. To compare them to one another, relative RFs should be
used \citep{1994A&A...283..129R,1996SoPh..164..169D}. These relative
responses are obtained by multiplying the standard, absolute RFs by
the corresponding model parameter. Relative RFs tell us how
sensitive one model parameter is compared to the others. For
instance, the relative RF to $\Delta\lambda_\mathrm{D}$ is much larger
than that to $\eta_0$ and that to $a$ (in particular three times as
large as the RFs to $\eta_0$ and twenty times larger than those to $a$
for Stokes $I$, in our sample M-E atmosphere). This means that a small
relative perturbation of $\Delta\lambda_\mathrm{D}$ changes the Stokes
profiles much more than the same relative perturbation of $\eta_0$ or
$a$. Consequently, $\Delta\lambda_\mathrm{D}$ should be better
determined by M-E inversion codes.

\subsection{Two-component model atmospheres}

Model atmospheres with two or more components are commonly used in the
analysis of observations. Any two-component model atmosphere is based on the
assumption that within the resolution element two different atmospheres
coexist, namely, one magnetic atmosphere filling a surface fraction $\alpha$,
and one non-magnetic atmosphere filling the remaining $(1-\alpha)$
fraction. $\alpha$ is called the magnetic filling factor. If
$\vec{I}_\mathrm{m}$ stands for the Stokes profile vector emerging from the
magnetic region and $\vec{I}_\mathrm{nm}$ for that of the non-magnetized
atmosphere, the observed Stokes vector can be written as
${\vec{I}}={\vec{I}_\mathrm{m} }\alpha+{\vec{I}_\mathrm{nm}}(1-\alpha)$.

Thus, according to Eq.~(\ref{eqresponse}), the RFs to $\alpha$
perturbations are given by
$\vec{I}_\mathrm{m}-\vec{I}_\mathrm{nm}$. Hence, the larger the
difference between the magnetic and the non-magnetic atmospheres, the
greater the sensitivity to $\alpha$. Since most of the difference is
the polarization signal itself, $Q_\mathrm{m}$, $U_\mathrm{m}$,
$V_\mathrm{m}$, when this signal is strong we can easily discern it
from the non-magnetic signal.

\subsection{The influence of smearing}

Spectral line smearing by macroturbulence is a well known effect that needs be
taken into account in the analysis of most observations except, perhaps, in
those with very high spatial resolution \citep{2000A&A...359..729A}. Besides
macroturbulence, instruments have finite-width profiles that produce smearing
of the observed Stokes profiles which become wider and with smaller
peaks. This smearing reduces the information on physical parameters
carried by the spectral line through convolution:
$\vec{I}_\mathrm{obs}=\vec{I}*F(\lambda)$, where * stands for the convolution
symbol and the scalar smearing profile, $F(\lambda)$, is convolved with all
the four Stokes parameters.

This loss of information through smearing is also translated into a loss of
sensitivity of spectral lines to the atmospheric quantities. In fact, since
the derivative of a convolution is equal to the convolution of
the derivative of one of the functions with the second one, response functions
become smeared as well:
\begin{equation}
\vec{R}_{\mathrm{obs},x}=\vec{R}_x*F(\lambda).
\end{equation}

\begin{figure*}[!ht]
  \centering
  \resizebox{0.47\hsize}{!}{\includegraphics{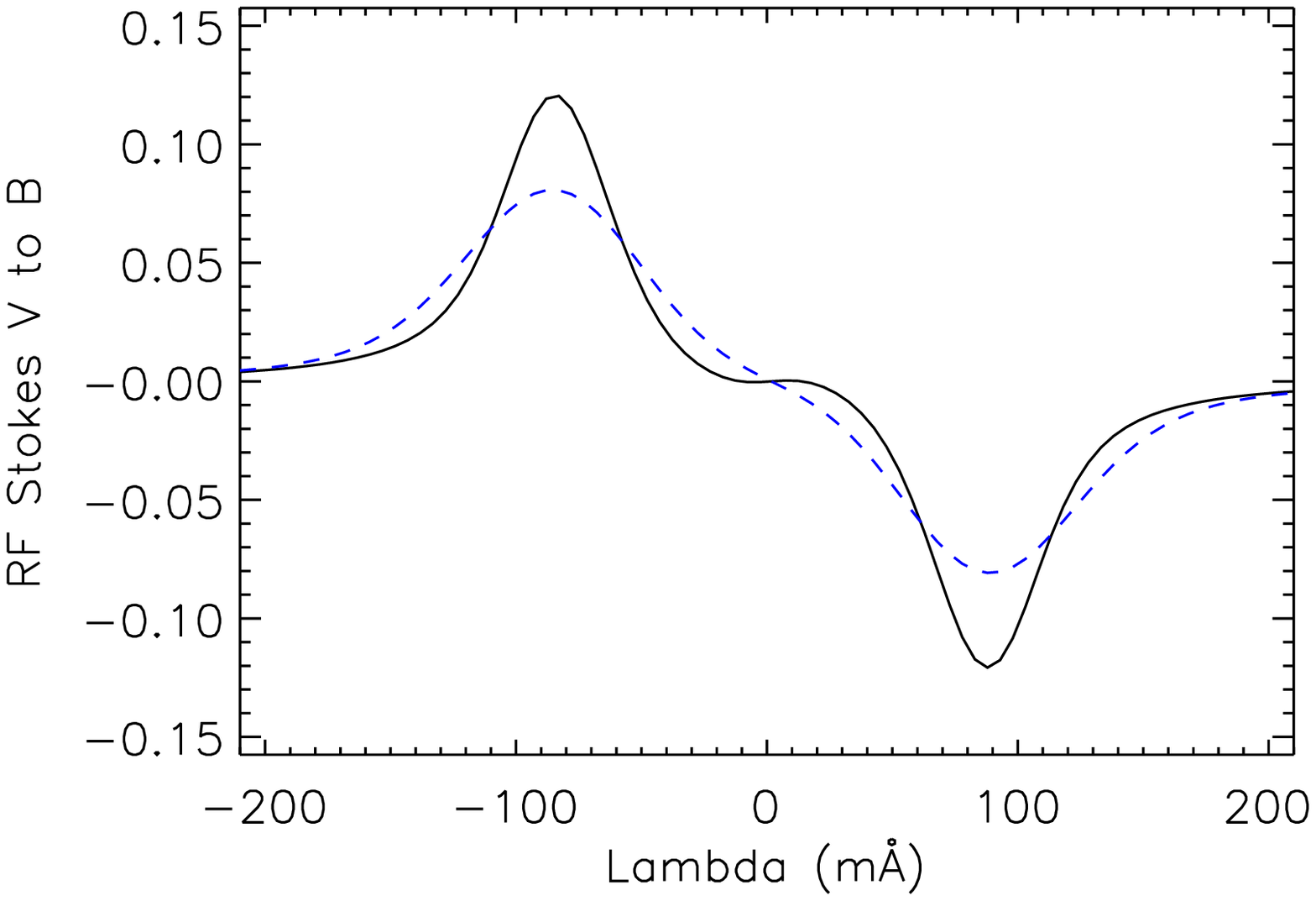}}
  \resizebox{0.47\hsize}{!}{\includegraphics{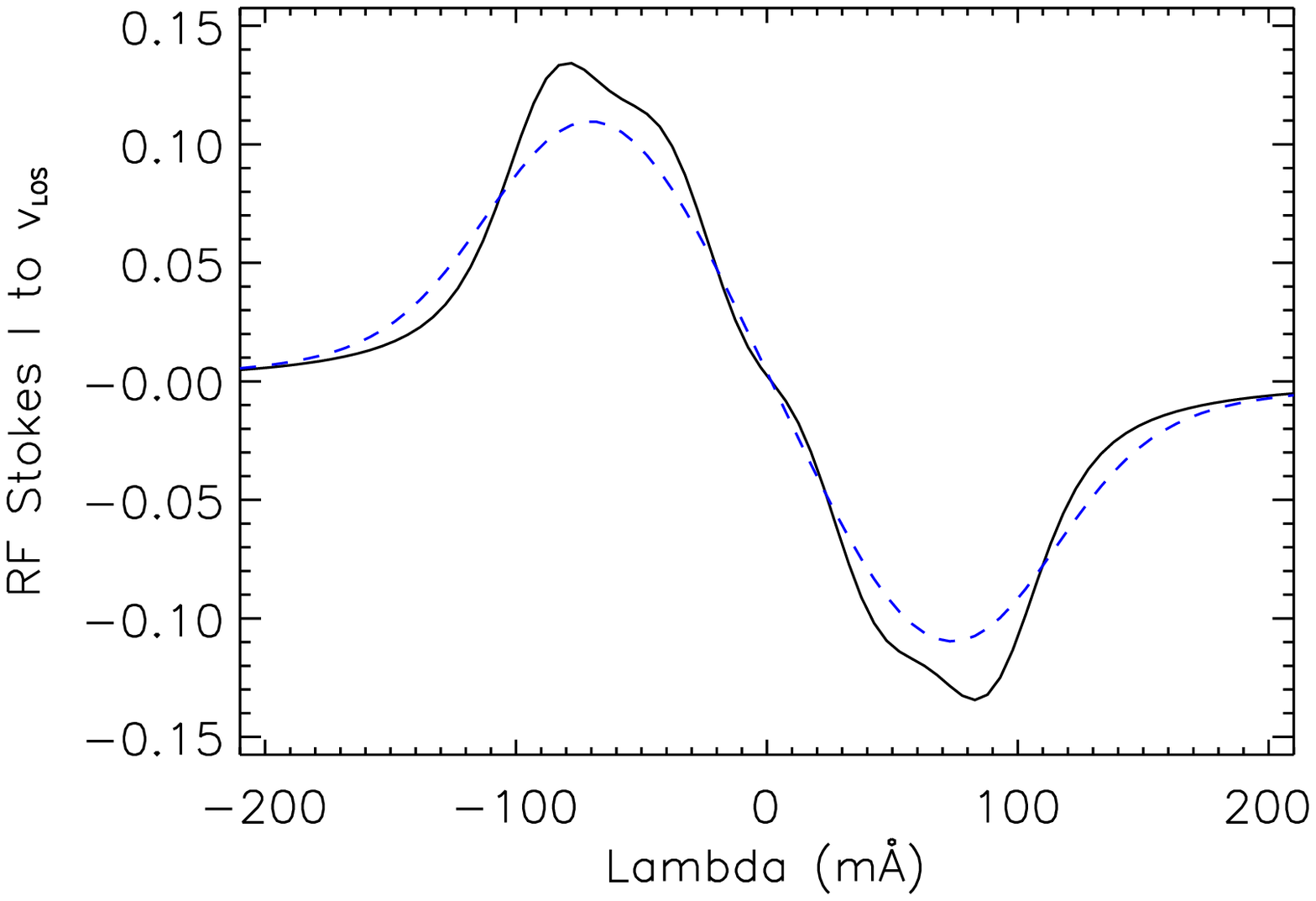}}
  \caption{Analytical M-E RFs of Stokes $V/I_{\mathrm{c}}$ to the
magnetic field strength (left panel) and of Stokes $I/I_{\mathrm{c}}$
to the LOS velocity (right panel) for the \ion{Fe}{i} line at 525.06
nm, with a magnetic field strength of 2000 G and field inclination and
azimuth of 45 degrees. The dashed lines correspond to the RFs
convolved with a Gaussian smearing profile of 60 m\AA ~of full width
at half maximum (FWHM). Solid lines correspond to the original
RFs. Units are in 10$^{-3}$ G$^{-1}$ (left) and [km/s]$^{-1}$
(right).}
  \label{Fig4rfs}
\end{figure*}

Fig.~\ref{Fig4rfs} shows the effect of RF smearing. The convolved RFs are
smoother and significant information is lost.

\subsection{The influence of noise}

Stokes profiles are affected by the noise intrinsic to the observational
process. Should the polarimetric signal be buried by noise, any algorithm one
could devise to determine atmospheric quantities would
fail. Therefore, our ability to infer accurate solar parameters depend
significantly on the signal-to-noise ratio of the observations.

Response functions can help in quantifying this effect. Since RFs
simply provide the modification of the Stokes profiles after a perturbation of
the physical quantities, if that modification is smaller than the noise level
it will be effectively undetectable. In other words, the size of RFs to
perturbations of a given quantity sets a threshold for the detection of a unit
of such a quantity: for instance, according to Fig.~\ref{Fig1rfs}, 1 Gauss will only be
detectable by a single wavelength sample if the noise is below 1.5$\cdot$10$^{-4}$
(continuum intensity is at 1); within the linear approximation\footnote{RFs
come from a linear perturbation analysis of the radiative transfer
equation}, 10 Gauss will be detectable with a noise below 1.5$\cdot$10$^{-3}$
and so on.

According to Eq.(~\ref{eqresponse}), a standard deviation in the Stokes signal
$\sigma$ will induce an error $\sigma_x$ per single wavelength sample given by:
\begin{equation}
\sigma=\sqrt {{R}^2_{1,x}+{R}^2_{2,x}+{R}^2_{3,x}+{R}^2_{4,x}} \sigma_x.
\end{equation}
Detectability should increase, of course, as the root of the wavelength sample
number, but the RFs allow an estimate of the expected
accuracy of our inferences.

The above estimates can be considered lower limits for the errors since model
parameters are not independent of each other and correlations may exist
between sensitivities such as those already reported between $\eta_0$,
$\Delta\lambda_\mathrm{D}$ and $a$.

\section{The usefulness of the RFs for vector magnetographs}

   \begin{figure*}[!ht]
   \centering
  \resizebox{0.9\hsize}{!}{\includegraphics[width=\textwidth]{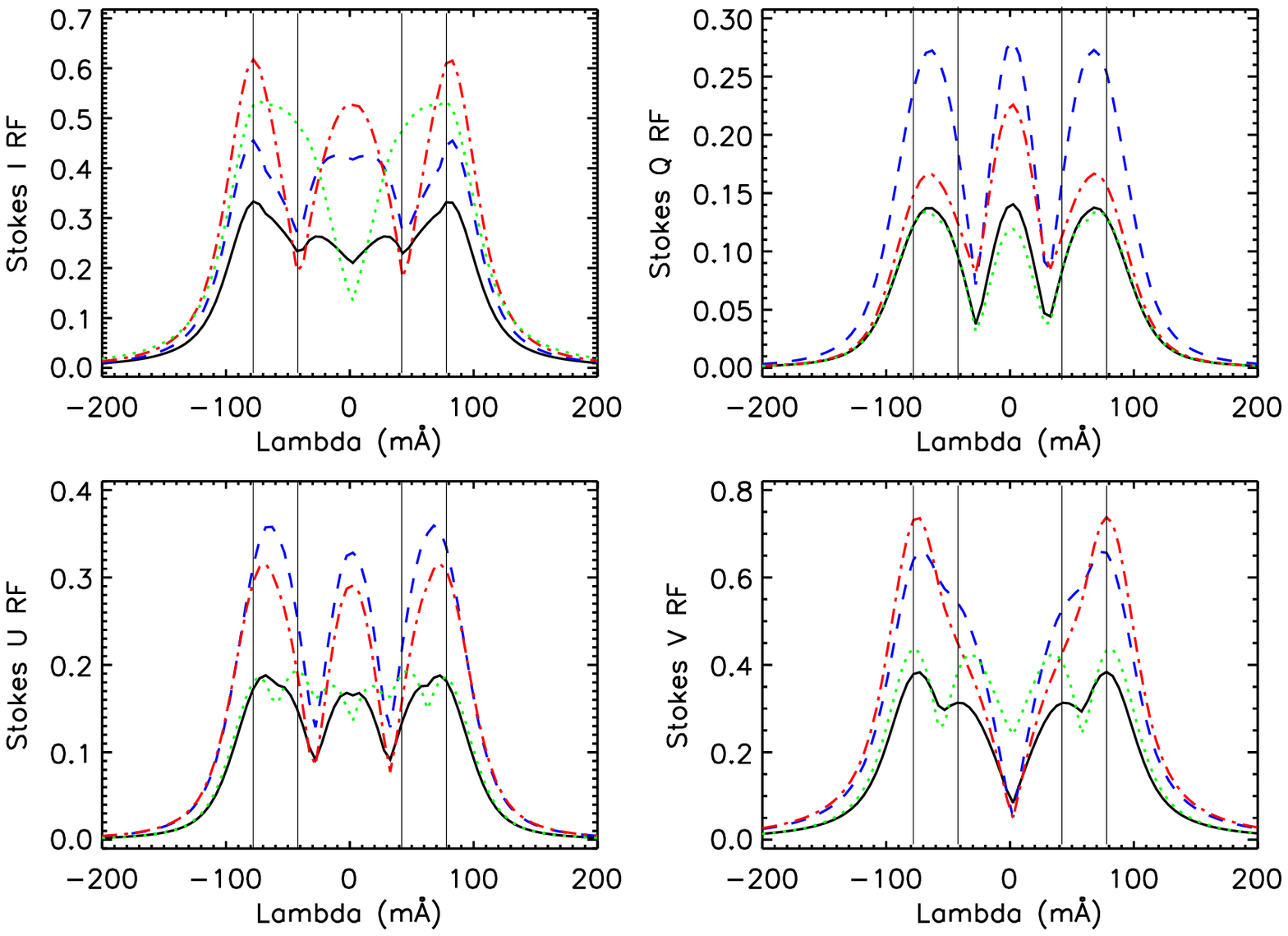}}
   \caption{Four different linear combinations of the Stokes vector RFs for
the IMaX line.  The plotted curves correspond to $\alpha_{1,2,3,4} = 1, 1, 1,
1$ (solid, black lines), $\alpha_{1,2,3,4} = 2, 2, 2 , 0.5$ (dashed, blue
lines), $\alpha_{1,2,3,4} = 3, 1, 1, 0.5$ (dashed-dotted, red lines), and
$\alpha_{1,2,3,4} = 0.5, 1, 1, 3$ (dotted, green lines). The light-grey,
vertical lines indicate a possible choice for wavelength sampling ($\pm 42,
78$ m\AA).}
  \label{Figall}
    \end{figure*}

\begin{figure}[!ht]
  \centering \resizebox{0.92\hsize}{!}{\includegraphics[width=3cm,trim=10mm
  100mm 0mm 0mm,clip]{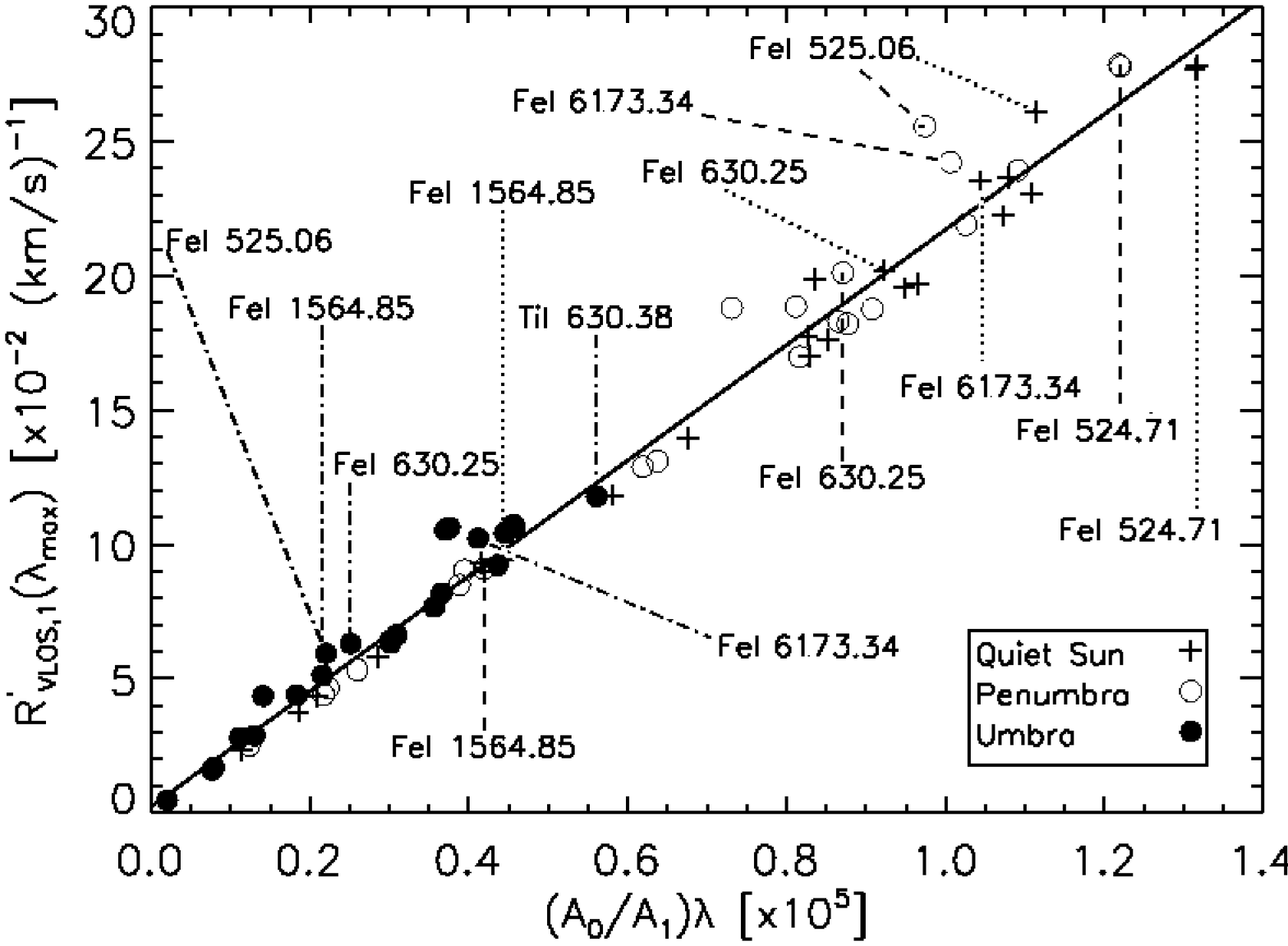}}
  \resizebox{0.89\hsize}{!}{\includegraphics[width=3cm,trim=13mm 107mm 0mm
  0mm,clip]{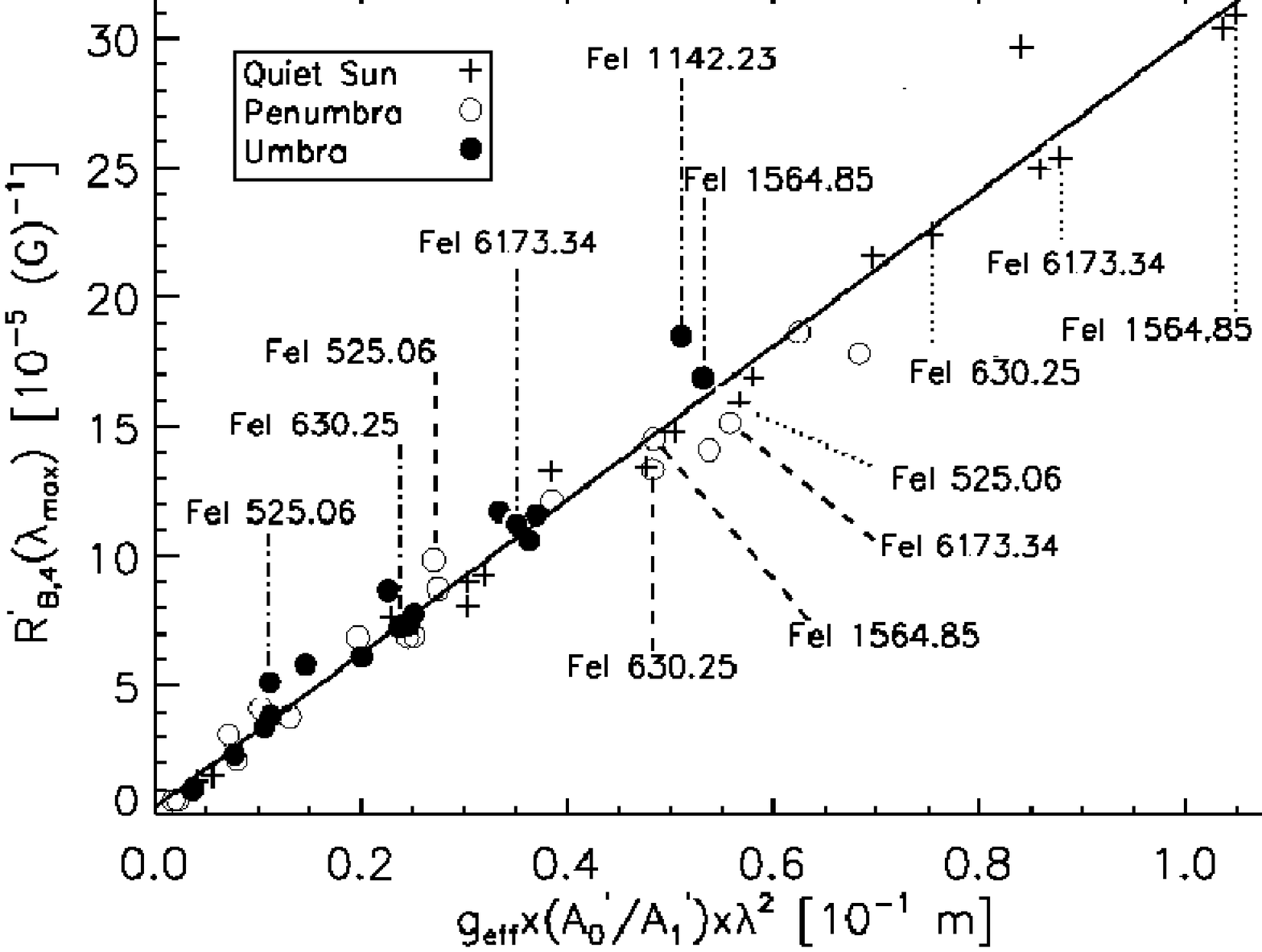}}
  \resizebox{0.92\hsize}{!}{\includegraphics[trim=10mm 100mm 0mm
  0mm,clip]{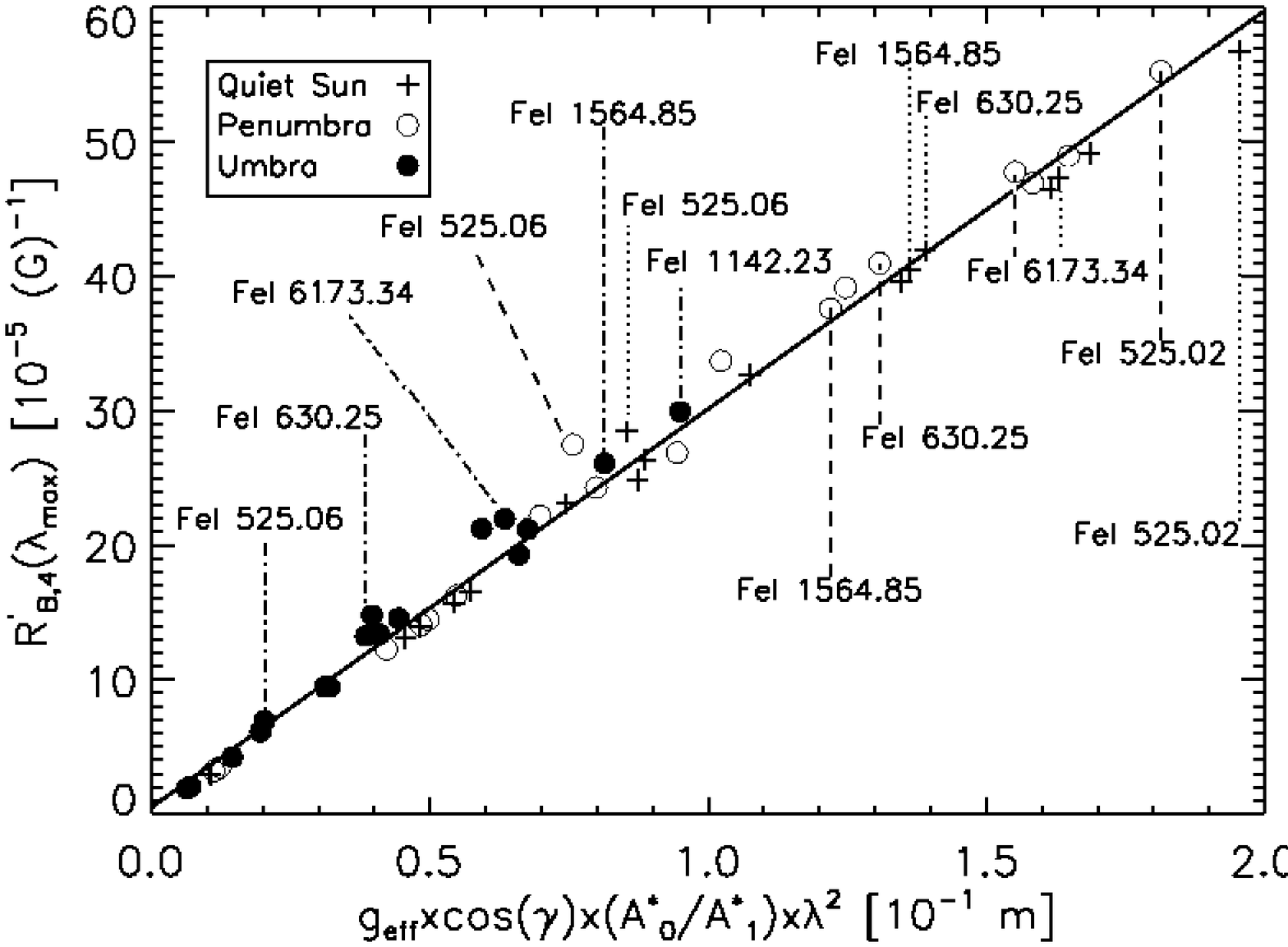}}
  \caption{Upper panel: Maximum value of the $\tau$-integrated
    RF to $v_\mathrm{LOS}$ for different lines as a function of the
    shape ratio multiplied by the central wavelength of the transition
    \citep[see][]{2005A&A...439..687C}. Middle panel: Maximum values
    of the integrated RF to $B$ for the same set of lines with
    g$_\mathrm{eff}$$\ne$0, as a function of the shape ratio
    multiplied by the squared central wavelength (strong field
    regime). Bottom panel: Maximum values of the integrated RF to $B$
    (weak field regime). The sensitivities have been evaluated in the
    quiet Sun (crosses), penumbral (circles) and hot umbral (filled
    circles) model atmospheres. Dotted, dashed, and dash-dotted lines
    mark specific transitions in the quiet sun, penumbral, and umbral
    models, respectively.}
  \label{cabfig}
\end{figure}

Modern vector magnetographs are not restricted to using one or two
wavelength samples as are the classical magnetographs. Instruments like
IMaX are devised to measure up to five wavelengths: one in the
continuum and four across the line profile. The choices of the
spectral line, of the number of samples and of the precise wavelength
for each of them are important issues that arise during the design
phase of the instrument. This section is aimed at illustrating how the
RFs can help such decisions.

Finding a suitable spectral line is crucial and can be achieved
through RFs on the simple phenomenological model by
\cite{2005A&A...439..687C} that allows establishing a ranking of
sensitivities to the different atmospheric parameters among the
various lines considered. The IMaX \ion{Fe}{i} line at 525.064 nm  
can be seen in Fig.~\ref{cabfig}. Data for this line have been
included in the original figure by \cite{2005A&A...439..687C}, where
it is identified as one of the most sensitive of the set to velocity
perturbations. It has a medium sensitivity to magnetic field strength
perturbations in both the strong and the weak field regimes.
However, it is not very sensitive to temperature (not shown) and,
hence, a good candidate for inferences in the various solar structures
avoiding thermodynamic trade-offs. The Helioseismic and Magnetic
Imager \citep[HMI;][]{HMI} and the Visible-light Imager and
Magnetograph \citep[VIM;][]{VIM}, two planned instruments for the {\em
Solar Dynamics Observatory}, NASA, and the {\em Solar Orbiter}, ESA,
missions, will use the \ion{Fe}{i} line at 617.334
nm. This spectral line is very well ranked in Fig.~\ref{cabfig} for
inference of both magnetic field strengths and LOS velocities.

A minimum number of wavelength samples is obtained by roughly doubling the
free parameters of the model: since an M-E model is made up of 
just ten parameters, a minimum of twenty observables (five wavelengths times
the four Stokes parameters) is needed. This is the choice for all the three
instruments mentioned above. Unfortunately, no purely objective means exists to
select the wavelengths for the samples. Nevertheless, RFs are a powerful tool
that help select those wavelengths that better suit our purposes. If one is
interested, for instance, in just the magnetic field strength and neglects the
other physical quantities, choosing those wavelengths where the RFs to $B$
reach local maxima would be appropriate. If the interest lies in several
physical quantities at the same time (e.g. on the three components of the
magnetic field and on the LOS velocity) we suggest the use of a linear
combination of regular RFs weighted according to the specific interests. 
Since RFs can be positive or negative, we propose the use of
absolute-valued RFs. Hence, consider
\begin{equation} 
\label{eqlinearcombination} 
{\cal R}_j = \sum_i \alpha_i \, |R_{j,i}|,
\end{equation}
where $j$ runs from 1 through 4, corresponding to the four Stokes parameters,
and index $i$ accounts for the physical parameters. Since the set of weights
$\alpha_i$ can be tailored at will, there is no single choice for samples but
an examination of ${\cal R}$ provides important hints for the selection. As
an example, Fig.~\ref{Figall} shows different linear combinations for the
IMaX line case.  If index $i$ runs from 1 through 4 standing for $B$,
$\gamma$, $\phi$, and $v_{\mathrm{LOS}}$, respectively, the plotted curves
correspond to $\alpha_{1,2,3,4} = 1, 1, 1, 1$ (solid, black lines),
$\alpha_{1,2,3,4} = 2, 2, 2 , 0.5$ (dashed, blue lines), $\alpha_{1,2,3,4} =
3, 1, 1, 0.5$ ( dashed-dotted, red lines), and $\alpha_{1,2,3,4} = 0.5, 1, 1,
3$ (dotted, green lines). The vertical lines indicate a possible
choice for wavelength sampling ($\pm 42, 78$ m\AA), selected mostly from the
properties of the Stokes $I$ and $V$ RFs since these two parameters usually
exhibit the largest signals in solar atmospheres.  While the most external
samples seem to be optimum, some other good choices for the inner
wavelengths are possible and up to the user.

\section{Conclusions}
\label{secconclu}

The many interesting features of analytic response functions have
been discussed in this paper by considering the specific case of an M-E model
atmosphere. Since an analytic solution for the radiative transfer equation is
available for this atmosphere, the sensitivities of spectral lines, as
described by RFs, can also be cast in an analytical form by simply taking partial
derivatives of such a solution with respect to the model parameters. The
analytic M-E solution has been thoroughly used in the past for
insight into radiative transfer physics and as a purely practical
diagnostic tool through the M-E inversion codes. Likewise, we have shown in
this paper that the analytic, M-E RFs are useful to better understand spectral
line formation and the behavior of Stokes profiles in different formation
conditions and also for practical recipes that can help in selecting
spectral lines for given purposes, in selecting wavelength samples, etc.

A summary of the various results obtained follows:

\begin{enumerate}
\item Response functions look homologous to each other, hence enabling 
qualitative, general discussions by considering a single spectral line in a 
specific model atmosphere. Here, we have targeted the IMaX, Fe {\sc i} line at 
525.60 nm in a M-E model representative of the quiet Sun thermodynamics (as
observed  by FTS) and with various vector magnetic fields and LOS velocities.

\item The sensitivities of spectral lines to the various 
parameters depend on wavelength: some samples are better suited to diagnose a 
given parameter; some wavelengths are even insensitive to another  
parameter. The RF extrema show trivially those wavelengths where sensitivity is 
maximum.

\item As expected in M-E conditions where no gradient of LOS 
velocity is present, RFs display clear wavelength symmetry properties. 
The RFs to magnetic field strength perturbations show similar parity 
as the Stokes profiles while the RFs to LOS velocity perturbations are of 
opposite parity.

\item Stokes $V$ sensitivities to $B$ perturbations are significant for very weak 
field strengths. This fact explains the reasonably accurate results of M-E 
inversions in this strength regime.

\item The shape of the RFs to LOS velocity perturbations does not depend on 
$v_{\rm LOS}$ except for the Doppler shift. Variations of sensitivity of the
Stokes $I$ and $V$ profiles are compensated: when information on $v_{\rm LOS}$
decreases in Stokes $I$ it increases in Stokes $V$, so that $v_{\rm LOS}$
remains well  inferred in any circumstance.

\item We understand the trade-offs often found in the inversion codes among M-E
thermodynamic  parameters: their corresponding RFs are very similar to each
other. Fortunately,  they are different from the other RFs and can
accurately infer vector  magnetic fields and LOS velocities. Among the
thermodynamic parameters, the relative sensitivity to $\Delta\lambda_D$ 
perturbations is larger than that to $\eta_0$ and $a$, hence enabling better 
inferences.

\item The magnetic filling factor $\alpha$ is better determined if there are 
significant differences between magnetic and non-magnetic atmospheres. When $B$ 
is large this result is natural; when $B$ is small, this result explains
that differences in the thermodynamics of both atmospheres can help in
inferring $\alpha$ properly.

\item Direct estimates of affordable noise levels can be directly obtained from 
RFs.

\item Response functions can also be used to select spectral lines for given 
purposes or for given measurements. Moreover, a suitable combination of RFs 
provides quantitive arguments for wavelength sample choice.
\end{enumerate}

\acknowledgements Interesting discussions with D. Cabrera Solana and L.R. Bellot 
Rubio are thanked. This work has been partially funded by Spanish Ministerio de
Educaci\'on y Ciencia through Project No. ESP2003-07735-C04-03 including a 
percentage from European FEDER funds.

\bibliographystyle{aa}

\begin{appendix}

\section{Explicit formulae}

The propagation matrix $\mathbf{K}$ of the RTE can be cast in the form
\citep[e.g.][]{2003insp.book.....D}
\begin{equation}
\mathbf{K}=\left( \begin{array}{cccc} \eta_I & \eta_Q & \eta_U
& \eta_V \\ \eta_Q & \eta_I & \rho_V & -\rho_U \\ \eta_U & -\rho_V &
\eta_I & \rho_Q \\ \eta_V & \rho_U & -\rho_Q & \eta_I \end{array}
\right),
\end{equation}
where
\label{acoff}
\begin{eqnarray}
\eta_I & = &
1+\frac{\eta_0}{2}\left[\phi_p\sin^2\gamma+\frac{\phi_b+\phi_r}{2}(1+\cos^2\gamma)\right],
\nonumber \\ \eta_Q & = &
\frac{\eta_0}{2}\left[\phi_p-\frac{\phi_b+\phi_r}{2}\right]\sin^2\gamma
\cos2\chi, \nonumber \\ \eta_U & = &
\frac{\eta_0}{2}\left[\phi_p-\frac{\phi_b+\phi_r}{2}\right]\sin^2\gamma
\sin2\chi, \nonumber \\ \eta_V & = &
\frac{\eta_0}{2}\left[\phi_r-\phi_b\right]\cos\gamma, \nonumber \\ \rho_Q & = &
\frac{\eta_0}{2}\left[\psi_p-\frac{\psi_b+\psi_r}{2}\right]\sin^2\gamma
\cos2\chi, \nonumber \\ \rho_U & = &
\frac{\eta_0}{2}\left[\psi_p-\frac{\psi_b+\psi_r}{2}\right]\sin^2\gamma
\sin2\chi, \nonumber \\ \rho_V & = &
\frac{\eta_0}{2}\left[\psi_r-\psi_b\right]\cos\gamma,
\end{eqnarray}
and $\phi_{p,b,r}$ and $\psi_{p,b,r}$ are the absorption and dispersion
  profiles, the $p,b,r$ indices stand for the $\pi$ and $\sigma$ components of
  a Zeeman multiplet, and $\eta_0$ is the ratio of line to continuum absorption
  coefficients.

$\phi_{p,b,r}$ and $\psi_{p,b,r}$ can be written as a sum of as many
absorption and dispersion profiles as the number of $p,b,r$ components as
follows:
\begin{eqnarray}
\phi_j & = & \displaystyle\sum_{M_l-M_u=j} S_{M_lM_u,j} H(a,\upsilon),
\nonumber \\
\psi_j & = & 2\displaystyle\sum_{M_l-M_u=j} S_{M_lM_u,j} F(a,\upsilon),
\end{eqnarray}
$S_{M_lM_u,j}$ being the strength of each component with
$j=-1,0,1$ corresponding to $b,p$ and $r$. $\upsilon$ stands for the
wavelength in Doppler units which follows
\begin{equation}
\upsilon = \frac{\lambda-\lambda_0}{\Delta \lambda_\mathrm{D}}+\frac{\Delta
\lambda_\mathrm{B}}{\Delta \lambda_\mathrm{D}}-\frac{\lambda_0
v_\mathrm{LOS}}{c \Delta \lambda_\mathrm{D}}.
\end{equation}

The wavelength shift of the different Zeeman components with respect to the
original position is given by
\begin{equation}
\Delta \lambda_\mathrm{B} = \frac{e \lambda_0^2 B}{4\pi mc^2}(g_lM_l-g_uM_u),
\end{equation}
where $l$ and $u$ stand for the lower and upper levels of the line transition,
$g$ for the level Land\'e factor, and $M$ for the magnetic level quantum
number.

The evaluation of RFs reduces to the derivatives of the Stokes
vector, $\vec{I}=(I,Q,U,V)$, with respect to the nine parameters,
$(B_0,B_1,\eta_0,B,\gamma,\chi,\Delta
\lambda_\mathrm{D},V_\mathrm{LOS},a)$. In order to easily show such
derivatives suppose a generic parameter $x$. Then,
\begin{eqnarray}
\frac{\partial{I}}{\partial{x}} & = & B_1 \mu \left( T_1
\frac{\partial{\ei}}{\partial{x}}+\ei \frac{\partial{T_1}}{\partial{x}} -
\Delta^{-1} \ei T_1 \frac{\partial{\Delta}}{\partial{x}}\right) \Delta^{-1},\\
\frac{\partial{Q}}{\partial{x}} & = & -B_1 \mu
\left(\frac{\partial{T_2}}{\partial{x}}+\frac{\partial{\rqq}}{\partial{x}}\Pi
+ \rqq \frac{\partial{\Pi}}{\partial{x}} - \Delta^{-1}
\frac{\partial{\Delta}}{\partial{x}} [T_2 + \rqq \Pi] \right) \Delta^{-1},  \nonumber\\
\frac{\partial{U}}{\partial{x}} & = & -B_1 \mu
\left(\frac{\partial{T_3}}{\partial{x}}+\frac{\partial{\ru}}{\partial{x}}\Pi +
\ru \frac{\partial{\Pi}}{\partial{x}} - \Delta^{-1}
\frac{\partial{\Delta}}{\partial{x}} [T_3 + \ru \Pi] \right) \Delta^{-1},  \nonumber\\
\frac{\partial{V}}{\partial{x}} & = & -B_1 \mu
\left(\frac{\partial{T_4}}{\partial{x}}+\frac{\partial{\rv}}{\partial{x}}\Pi +
\rv \frac{\partial{\Pi}}{\partial{x}} - \Delta^{-1}
\frac{\partial{\Delta}}{\partial{x}} [T_4 + \rv \Pi] \right) \Delta^{-1},  \nonumber
\end{eqnarray}
where for simplicity
\begin{eqnarray}
T_1 & = & \ei^2+\rqq^2+\ru^2+\rv^2, \nonumber \\
T_2 & = & \ei^2 \eq+\ei(\ev\ru-\eu\rv), \nonumber \\
T_3 & = & \ei^2 \eu+\ei(\eq\rv-\ev\rqq), \nonumber\\
T_4 & = & \ei^2 \ev+\ei(\eu\rqq-\eq\ru), \nonumber \\
T_5 & = & \ei^2-\eq^2-\eu^2-\ev^2+\rqq^2+\ru^2+\rv^2. 
\end{eqnarray}

$\Delta$ and $\Pi$ are defined in Eqs.~(4) and (5), respectively. Their
derivatives are thus given by
\begin{eqnarray}
\frac{\partial{\Delta}}{\partial{x}} & = & 2\ei
\frac{\partial{\ei}}{\partial{x}} T_5 + \ei^2\frac{\partial{T_5}}{\partial{x}}
- 2 \Pi\frac{\partial{\Pi}}{\partial{x}},  \\
\frac{\partial{\Pi}}{\partial{x}} & = & \eq
\frac{\partial{\rqq}}{\partial{x}}+ \frac{\partial{\eq}}{\partial{x}}\rqq+
\eu \frac{\partial{\ru}}{\partial{x}}+ \frac{\partial{\eu}}{\partial{a}} \ru+
\ev \frac{\partial{\rv}}{\partial{x}}+ \frac{\partial{\ev}}{\partial{x}} \rv. \nonumber
\end{eqnarray}

The derivatives of $T_1,...,T_5$ are given by
\begin{eqnarray}
\frac{\partial{T_1}}{\partial{x}} & = & 2 \left(\ei
\frac{\partial{\ei}}{\partial{x}}+ \rqq \frac{\partial{\rqq}}{\partial{x}}+\ru
\frac{\partial{\ru}}{\partial{x}}+ \rv
\frac{\partial{\rv}}{\partial{x}}\right), \nonumber \\
\frac{\partial{T_2}}{\partial{x}} & = &
2\ei\frac{\partial{\ei}}{\partial{x}}\eq +
\ei^2\frac{\partial{\eq}}{\partial{x}}+\frac{\partial{\ei}}{\partial{x}}(
\ev\ru-\eu\rv ) + \nonumber \\ & & \ei \left(\frac{\partial{\ev}}{\partial{x}}
\ru+ \ev \frac{\partial{\ru}}{\partial{x}}-\frac{\partial{\eu}}{\partial{x}}
\rv- \eu \frac{\partial{\rv}}{\partial{x}} \right) , \nonumber\\
\frac{\partial{T_3}}{\partial{x}} & = &
2\ei\frac{\partial{\ei}}{\partial{x}}\eu +
\ei^2\frac{\partial{\eu}}{\partial{x}}+\frac{\partial{\ei}}{\partial{x}}(
\eq\rv-\ev\rqq ) + \nonumber \\ & & \ei
\left(\frac{\partial{\eq}}{\partial{x}} \rv+ \eq
\frac{\partial{\rv}}{\partial{x}}-\frac{\partial{\ev}}{\partial{x}} \rqq- \ev
\frac{\partial{\rqq}}{\partial{x}} \right) , \nonumber\\
\frac{\partial{T_4}}{\partial{x}} & = &
2\ei\frac{\partial{\ei}}{\partial{x}}\ev +
\ei^2\frac{\partial{\ev}}{\partial{x}}+\frac{\partial{\ei}}{\partial{x}}(
\eu\rqq-\eq\ru ) + \nonumber \\ & & \ei
\left(\frac{\partial{\eu}}{\partial{x}} \rqq+ \eu
\frac{\partial{\rqq}}{\partial{x}}-\frac{\partial{\eq}}{\partial{x}} \ru- \eq
\frac{\partial{\ru}}{\partial{x}} \right) , \nonumber\\
\frac{\partial{T_5}}{\partial{x}} & = & 2 \left( \ei
\frac{\partial{\ei}}{\partial{x}}- \eq \frac{\partial{\eq}}{\partial{x}}-\eu
\frac{\partial{\eu}}{\partial{x}}- \ev \frac{\partial{\ev}}{\partial{x}} +
\rqq \frac{\partial{\rqq}}{\partial{x}}+ \right. \nonumber \\ 
& & \left. \ru \frac{\partial{\ru}}{\partial{x}}+ \rv \frac{\partial{\rv}}{\partial{x}} \right).
\end{eqnarray}

The derivatives with respect to $\eta_0$ can be easily calculated from
Eq.~(A.\/2.)
\begin{eqnarray}
\frac{\partial{\ei}}{\partial{\eta_0}} & = & \frac{(\ei-1)}{\eta_0}, \nonumber \\
\frac{\partial{\eta_{Q,U,V}}}{\partial{\eta_0}} & = & \frac{\eta_{Q,U,V}}{\eta_0},
\nonumber \\
\frac{\partial{\rho_{Q,U,V}}}{\partial{\eta_0}} & = & \frac{\rho_{Q,U,V}}{\eta_0}.
\end{eqnarray}

 The derivatives with respect to
$\gamma$ and $\psi$ are 
\begin{eqnarray}
\frac{\partial{\ei}}{\partial{\chi}} & = & 0, \,\,\,\,\,\,
\frac{\partial{\ev}}{\partial{\chi}} = 0, \,\,\,\,\,\,
\frac{\partial{\rv}}{\partial{\chi}} = 0, \nonumber \\
\frac{\partial{\eq}}{\partial{\chi}} & = & -2 \eq \tan 2\chi, \nonumber \\
\frac{\partial{\eu}}{\partial{\chi}} & = & 2 \eu \cot 2\chi, \nonumber \\
\frac{\partial{\rqq}}{\partial{\chi}} & = & -2 \rqq \tan 2\chi, \nonumber \\
\frac{\partial{\ru}}{\partial{\chi}} & = & 2 \ru \cot 2\chi, \nonumber \\
\frac{\partial{\ei}}{\partial{\gamma}} & = & \frac{\eta_0}{2}\left[\phi_p-\frac{\phi_b+\phi_r}{2}\right]\sin2\gamma,
\nonumber \\
\frac{\partial{\eq}}{\partial{\gamma}} & = & \frac{\eta_0}{2}\left[\phi_p-\frac{\phi_b+\phi_r}{2}\right]\sin2\gamma\cos2\chi,
 \\
\frac{\partial{\eu}}{\partial{\gamma}} & = & \frac{\eta_0}{2}\left[\phi_p-\frac{\phi_b+\phi_r}{2}\right]\sin2\gamma\sin2\chi,
\nonumber \\
\frac{\partial{\ev}}{\partial{\gamma}} & = & -\ev\tan\gamma,  \nonumber \\
\frac{\partial{\rqq}}{\partial{\gamma}} & = & \frac{\eta_0}{2}\left[\psi_p-\frac{\psi_b+\psi_r}{2}\right]\sin2\gamma\cos2\chi,
\nonumber \\
\frac{\partial{\ru}}{\partial{\gamma}} & = & \frac{\eta_0}{2}\left[\psi_p-\frac{\psi_b+\psi_r}{2}\right]\sin2\gamma\sin2\chi,
\nonumber \\
\frac{\partial{\rv}}{\partial{\gamma}} & = & -\rv\tan\gamma. \nonumber 
\end{eqnarray}

The derivatives with respect to the other parameters imply the
derivatives of the absorption and dispersion profiles and these lead
us to obtain the derivatives of the Voigt and Voigt-Faraday
functions (as defined by Landi degl'Innocenti, 1976):
\begin{eqnarray}
\frac{\partial{\phi_j}}{\partial{x}} & = & \displaystyle\sum_{M_l-M_u=j} S_{M_lM_u,j}
\frac{\partial{H(a,\upsilon)}}{\partial{x}}, \nonumber \\
\frac{\partial{\psi_j}}{\partial{x}} & = & 2\displaystyle\sum_{M_l-M_u=j} S_{M_lM_u,j}
\frac{\partial{F(a,\upsilon)}}{\partial{x}}.
\end{eqnarray}

By using the chain rule and the derivatives of $H(a,\upsilon)$ and
$F(a,\upsilon)$ with to respect $a$ and $\upsilon$,
\begin{eqnarray}
\frac{\partial{H(a,\upsilon)}}{\partial{a}} & = & -2
\frac{\partial{F(a,\upsilon)}}{\partial{\upsilon}}, \nonumber \\
\frac{\partial{F(a,\upsilon)}}{\partial{a}} & = & \frac{1}{2}
\frac{\partial{H(a,\upsilon)}}{\partial{\upsilon}}, \nonumber \\
\frac{\partial{H(a,\upsilon)}}{\partial{\upsilon}} & = &
4aF(a,\upsilon)-2\upsilon H(a,\upsilon), \nonumber \\
\frac{\partial{F(a,\upsilon)}}{\partial{\upsilon}} & = &
\frac{1}{\sqrt{\pi}}-aH(a,\upsilon)-2\upsilon F(a,\upsilon),
\end{eqnarray}
we find
\begin{eqnarray}
\frac{\partial{H(a,\upsilon),F(a,\upsilon)}}{\partial{B}} & = &
\frac{\partial{H(a,\upsilon),F(a,\upsilon)}}{\partial{\upsilon}}\frac{\Delta
\lambda_{i_j}}{\Delta \lambda_\mathrm{D}} \frac{1}{B}, \nonumber \\
\frac{\partial{H(a,\upsilon),F(a,\upsilon)}}{\partial{v_\mathrm{LOS}}} & = &
\frac{\partial{H(a,\upsilon),F(a,\upsilon)}}{\partial{\upsilon}}\frac{-\lambda_0}{c
\Delta \lambda_\mathrm{D}}, \nonumber  \\
\frac{\partial{H(a,\upsilon),F(a,\upsilon)}}{\partial{\Delta\lambda_\mathrm{D}}}
& = & 
\frac{\partial{H(a,\upsilon),F(a,\upsilon)}}{\partial{\upsilon}}
\frac{-\upsilon}{\Delta\lambda_\mathrm{D}}.
\end{eqnarray}

\end{appendix}

\end{document}